\documentclass[twocolumn,pra,superscriptaddress]{revtex4-1}

\usepackage{epsfig}
\usepackage{graphicx}
\usepackage{amsmath}
\usepackage{amssymb}
\usepackage{epstopdf}
\usepackage{xcolor}
\usepackage{subfigure}
\usepackage{bm}
\usepackage{dsfont}
\usepackage{hyperref}
\usepackage{braket}
\usepackage{todonotes}
\usepackage{comment}

\begin{document}

\title{Decomposition of split-step quantum walks for simulating\\ Majorana modes and edge states}
\author{Wei-Wei Zhang}
\affiliation{Hefei National Laboratory for Physical Sciences at Microscale and Department of Modern Physics, University of Science and Technology of China, Hefei, Anhui 230026, China}
\affiliation{Institute for Quantum Science and Technology, and Department of Physics and Astronomy, University of Calgary, Canada, T2N~1N4}
\affiliation{State Key Laboratory of Networking and Switching Technology, Beijing University of Posts and Telecommunications, Beijing, 100876, China}

\author{Sandeep K.~Goyal}
\affiliation{Institute for Quantum Science and Technology, and Department of Physics and Astronomy, University of Calgary, Canada, T2N~1N4}
\affiliation{Department of Physical Sciences, Indian Institute of Science Education and Research Mohali, Sector 81, SAS Nagar, Punjab, 140306, India}
\email{skgoyal@iisermohali.ac.in}

\author{Christoph Simon}
\affiliation{Institute for Quantum Science and Technology, and Department of Physics and Astronomy, University of Calgary, Canada, T2N~1N4}

\author{Barry C.~Sanders}
\affiliation{Hefei National Laboratory for Physical Sciences at Microscale and Department of Modern Physics, University of Science and Technology of China, Hefei, Anhui 230026, China}
\affiliation{Institute for Quantum Science and Technology, and Department of Physics and Astronomy, University of Calgary, Canada, T2N~1N4}
\affiliation{Shanghai Branch, CAS Center for Excellence and Synergetic Innovation Center in Quantum Information and Quantum Physics, University of Science and Technology of China, Shanghai 201315, China}
\affiliation{Program in Quantum Information Science, Canadian Institute for Advanced Research, Toronto, Ontario M5G~1Z8, Canada}

\begin{abstract}
  We construct a decomposition procedure for converting split-step quantum walks into ordinary quantum walks with alternating coins, and we show that this decomposition enables a feasible linear optical realization of split-step quantum walks by eliminating quantum-control requirements.  As salient applications, we show how our scheme will simulate Majorana modes and edge states.

\end{abstract}

\maketitle

\section{Introduction}

Topologically ordered quantum states demonstrate many interesting properties such as fractional statistics, spin liquids, and robust ground-state degeneracy, which are the basis of topological and fault-tolerant quantum computation~\cite{Klitzing1980,Thouless1982,Laughlin1983,Wen1995,Stormer1999,Ryu2002,Kane2005,Kane2005a,Hsieh2008,Hasan2010,Bernevig2006,Nayak2008,Kitaev2003,Kitaev2006}. 
One-dimensional discrete time quantum walks and their one- and two-dimensional generalizations called split-step quantum walk (SSQW) exhibit a rich class of topological phases and exotic phases such as Majorana modes and edge states~\cite{Kitagawa2010,Obuse2011,Asboth2012,Rakovszky2015,Cedzich2015,Obuse2015,Groh2016,Lam2016}. We present a procedure to decompose one- and two-dimensional SSQWs into ordinary one-dimensional quantum walks (OQWs) with alternating coins. Using this decomposition we propose simple implementation schemes to realize one- and two-dimensional SSQWs in the linear optical setup.

The interface of two distinct topological phases can host topologically protected bounded states such as Majorana modes and edge states~\cite{Ryu2002,Kitagawa2010,Wilczek2014,Lam2016}.  Majorana modes~\cite{Wilczek2014} are quasiparticles which are their own antiparticles and the edge states are the low energy conducting states which exist on the surface (or the edges) of an insulating material~\cite{Tamm1932,Shockley1939,Hasan2010}. Edge states have been used to understand topological insulators and the Hawking radiations in black holes~\cite{Balachandran1997,Corichi1999}. Discrete time quantum walks provide controllable platforms to simulate and manipulate these exotic phases~\cite{Kitagawa2010,Obuse2011,Asboth2012,Rakovszky2015,Cedzich2015,Obuse2015,Groh2016}.

The OQW and the SSQW exhibit a large class of topological phases, the  Majorana modes and the edge states~\cite{Kitagawa2010,Asboth2012,Lam2016,Asboth2012}. Despite these advantages, the two-dimensional SSQW has never been implemented, whereas the realization of the one-dimensional SSQW was reported only in Ref.~\cite{Kitagawa2012}. This is partly due to the difficulty in implementing quantum walks in more than one dimension in a controllable way and partly due to an inadequate understanding of the SSQW.

In this article, we 
show that in spite of having very different propagators, the one- and the two-dimensional SSQWs and the OQW  are closely related; each step of the one-dimensional SSQW can be decomposed into two steps of the OQW with alternating coin-flip operations. Similarly, every step of the two-dimensional SSQW can be decomposed into two steps of one-dimensional SSQW performed over two independent degrees of freedom in sequence with the same coin.

The decomposition of the SSQW in terms of the OQW presented here yields a direct relation between the Hamiltonian of the OQW and the Hamiltonian for the SSQW. This decomposition shows that the SSQW can be thought of as a special case of alternate quantum walks~\cite{DiFranco2011,DiFranco2011a,Roldan2013}. It also paves the way to simulate more complicated Hamiltonians using only the OQW. Furthermore,
it enables simple schemes to implement complicated quantum walks on any accessible systems. Using our decompositions we present implementation schemes to realize  one- and two-dimensional SSQWs using a linear optical setup.

In our scheme, the one-dimensional quantum walk is performed over the orbital angular momentum (OAM) states of a single photon (or a light pulse) whereas we use the time-bins along with the OAM of light to realize two-dimensional SSQWs. The polarization of light serves as the coin in our scheme. The proposed setup  requires a simple combination of wave plates, $q$-plate, polarizing beamsplitters, and mirrors, and a  ring interferometer is used to implement progressive steps in the walk. 

Since both classical light pulses as well as single photons can possess the OAM and the time-bin degrees of freedom (DoFs), and the proposed setup for the implementation consists of only linear optical elements, our scheme works  equally well for single photons and classical light pulses. Simulating quantum protocols with classical light offers advantages over the single photons such as noninvasive and real-time measurements, which is not possible with single photons~\cite{Bouwmeester1999,Knight2003,Goyal2013}. Furthermore, classical light is robust against losses and easy to produce. 
Our scheme is capable of simulating the Majorana modes and the edge states with classical light, and the setup size does not increase with increasing number of steps in the walk, as was the case in the earlier implementation~\cite{Kitagawa2012}. This is the first scheme where the realization of such exotic modes in two dimensions is addressed.

The article is organized as follows: In Sec.~\ref{Sec:Background} we describe the one- and two-dimensional SSQWs, and the topological properties of the underlying Hamiltonians. Section~\ref{Sec:Results} deals with the decomposition of SSQWs into OQWs. In Sec.~\ref{Sec:Implementation} we present our implementation schemes and the methods to simulate the Majorana modes and the edge states. We conclude in Sec.~\ref{Sec:Conclusion}. 

\section{Background}\label{Sec:Background}
In this section, we present the relevant background of the OQW and SSQW, and the topological nature of these quantum walks. 

\subsection{Ordinary and split-step quantum walk}\label{Subsec:QW}

We start with the OQW where
the coin-flip operator $C_\theta$ and the conditional propagator $S$ are
\begin{align}
C_\theta &\equiv C(\theta) = \cos\theta\mathds{1} -\text{i} \sin\theta \sigma_y,\label{Eq:Coin}\\
S& = F\otimes\left| \uparrow \right \rangle \left \langle \uparrow\right |+F^{\dagger}\otimes\left|\downarrow\right\rangle\left\langle\downarrow\right|.\label{Eq:S}
\end{align}
Here, \{$\left|  \uparrow  \right\rangle ,\left|  \downarrow  \right\rangle$\} are the two orthogonal states in the coin space,  $\{\left| x \right\rangle , x \in \mathds{Z}\}$ are the position states of the walker such that $F =\sum_{x}\ket{x+1}\bra{ x}$ is the forward propagator, and $\sigma_y$ is the Pauli spin matrix along the $y$ axis.
The coin parameter $\theta \in (-\pi,\pi]$.

Repeated action of the propagator $Z(\theta)$,
\begin{align}\label{Eq:Z}
Z(\theta) &= (\mathds{1}\otimes C_\theta)S,
\end{align}
on the states of the walker results in the quantum walk evolution. Note that the operator $\bar{Z}(\theta) = S(\mathds{1}\otimes C_\theta)$ also yields the same quantum walk dynamics as the propagator $Z(\theta)$ for different initial states which are related to each other by the unitary operator $S$. Thus, the two operators, $Z(\theta)$ and $\bar{Z}(\theta)$, are equivalent. We shall call it the cyclic property of the quantum walk propagator.

In a one-dimensional SSQW~\cite{Kitagawa2010} the conditional propagator $S$~\eqref{Eq:S} is divided into the left propagator $T_-$ and the right propagator $T_+$ which are separated by a coin-flip operation $C_\theta$. Thus, the new quantum walk propagator reads
\begin{equation}
  Z_{\rm ss}(\theta_1,\theta_2) = (\mathds{1}\otimes C_{\theta_1}) T_-(\mathds{1}\otimes C_{\theta_2}) T_+ \equiv (\mathds{1}\otimes C_{\theta_1}) T_{\theta_2},\label{Eq:Zss}
\end{equation}
where
\begin{align}
  T_+ &= F\otimes \ket{\uparrow}\bra{\uparrow} + \mathds{1}\otimes \ket{\downarrow}\bra{\downarrow},\label{Eq:Tp}\\
  T_- &=  \mathds{1}\otimes \ket{\uparrow}\bra{\uparrow} + F^\dagger\otimes \ket{\downarrow}\bra{\downarrow},\label{Eq:Tm}\\
  T_{\theta_2} &=T_-(\mathds{1}\otimes C_{\theta_2}) T_+.\label{Eq:T-theta}
\end{align}

\begin{figure}
  \includegraphics[width=0.15\textwidth]{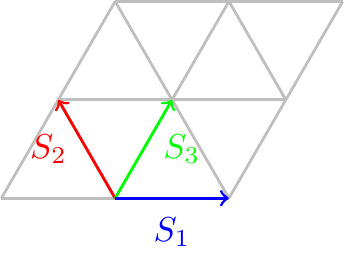}
  \caption{Schematics representation of two-dimensional SSQW on a triangular lattice. Here we show the directions of the conditional propagators $S_1$ and $S_2$  defined in Eqs.~(\ref{Eq:2DSS-T1}) and~(\ref{Eq:2DSS-T2}), and $S_3 = S_1S_2$. }
  \label{Fig:2DSST}
\end{figure}

The propagator $ Z_{\rm 2d}(\theta_1,\theta_2)$ for the two-dimensional analog of a SSQW on a triangular lattice consists of three conditional propagators $S_i$ applied in series, separated by the coin operations $\mathds{C}_{\theta}$~\cite{Kitagawa2010},
\begin{equation}
  Z_{\rm 2D}(\theta_1,\theta_2) = S_3\mathds{C}_{\theta_1}S_2\mathds{C}_{\theta_2}S_1 \mathds{C}_{\theta_1},\label{Eq:2DSS-U}
\end{equation}
where (see Fig.~\ref{Fig:2DSST})
\begin{align}
  S_1 &= (F\otimes \mathds{1}) \otimes \Ket{\uparrow}\Bra{\uparrow} + (F^\dagger \otimes \mathds{1}) \otimes \Ket{\downarrow}\Bra{\downarrow},\label{Eq:2DSS-T1}\\
  S_2 &= ( \mathds{1}\otimes F) \otimes \Ket{\uparrow}\Bra{\uparrow} + (\mathds{1}\otimes F^\dagger) \otimes \Ket{\downarrow}\Bra{\downarrow}\label{Eq:2DSS-T2}
\end{align}
are the conditional propagators on the two principal axes in the triangular lattice and $S_3 = S_1S_2$. The coin operator $\mathds{C}_{\theta_i} = \mathds{1}\otimes\mathds{1}\otimes C_{\theta_i}$.

\subsection{Topological phases in quantum walk}\label{Subsec:Topology}

The OQW~\eqref{Eq:Z} exhibits two distinct topological phases which can be characterized by the sign of the parameter $\theta$~\cite{Kitagawa2010,Asboth2012}. The interface of these two topological phases supports two bound states corresponding to (quasi-) energy $E = 0,\pi$~\cite{Kitagawa2010,Asboth2012,Lam2016}. The underlying Hamiltonian of this dynamics possesses the particle-hole symmetry which implies that creating a particle with energy $E$ is equivalent to annihilating a hole with energy $-E$. Since the two bound states in the quantum walk satisfy $ E = -E$, the creation and the annihilation operators for these states are the same. Thus, the corresponding modes are the Majorana modes~\cite{Lam2016}.

Both one- and two-dimensional SSQWs exhibit topological phases in the parameter space of $\theta_1,\theta_2$. 
Moreover, choosing site-dependent $\theta_2$ while keeping $\theta_1$ uniform over the lattice can result in a boundary such that we observe different topological phases on both sides of the boundary. Such boundaries support topologically protected bound states in the one-dimensional SSQW and edge states in the two-dimensional case~\cite{Kitagawa2010}.

Although SSQWs are known to simulate a large class of topological phases, schemes to implement these quantum walks in a controllable manner are not known. In the next section, we present a procedure for converting the SSQW into the OQW. 
\section{Decomposing SSQW}\label{Sec:Results}

In this section, we present the decomposition procedure for a single step of the  one- and the two-dimensional SSQWs in terms of OQWs. First, we decompose each step of the one-dimensional SSQW into two steps of an OQW with different coin operators (Sec.~\ref{Subsec:decomposition:1d}), and then we decompose the two-dimensional SSQW in terms of two one-dimensional SSQWs being performed on two different degrees of freedom or two different lattices (Sec.~\ref{Subsec:decomposition:2d}).

\subsection{Decomposing one-dimensional SSQW}\label{Subsec:decomposition:1d}

Here, we show that a single step of the one-dimensional SSQW is isomorphic to two steps of the  OQW. This isomorphism can be established easily by replacing the left and right propagators $T_-$ and $T_+$ in Eqs.~\eqref{Eq:Tp} and~\eqref{Eq:Tm} by $T_-^2$ and $T_+^2$.
Thus, the new one-dimensional SSQW propagator reads
\begin{equation}
  \tilde{Z}_{\text{ss}}(\theta_1,\theta_2) = (\mathds{1}\otimes C_{\theta_1}) {T}_-^2(\mathds{1}\otimes C_{\theta_2}) {T}_+^2.\label{Eq:Z2ss}
\end{equation}
Qualitatively, there is no difference between the propagator $Z_{\text{ss}}(\theta_1,\theta_2)$~\eqref{Eq:Zss} and the propagator $\tilde{Z}_{\text{ss}}(\theta_1,\theta_2)$~\eqref{Eq:Z2ss}. The only difference is in the former one, in which the walker jumps on the neighboring sites and in the latter one the walker skips one site in every jump.

Since the one-dimensional SSQW is translation invariant, we can write
\begin{align}
  \tilde{Z}_{\text{ss}}(\theta_1,\theta_2) &= \overrightarrow{T}\tilde{Z}_{\text{ss}}(\theta_1,\theta_2)\overleftarrow{T}\nonumber\\
                                           &= (\mathds{1}\otimes C_{\theta_1}) \overrightarrow{T}{T}_-^2(\mathds{1}\otimes C_{\theta_2}) \overleftarrow{T}{T}_+^2\nonumber\\
  &= Z(\theta_2) Z(\theta_1).\label{Eq:Zss-ZZ}
\end{align}
Here the coin-independent translation operators $\overrightarrow{T} = \overleftarrow{T}^\dagger= F \otimes \mathds{1}$ commute with $\tilde{T}_\pm$ operators. The operator $S = \overrightarrow{T}{T}_-^2 = \overleftarrow{T}{T}_+^2$ is the conditional shift operator~\eqref{Eq:S}.
Thus, the one-dimensional SSQW operator can be decomposed into two steps of the OQW propagators with alternating coin operators $C_{\theta_1}$ and $C_{\theta_2}$. In other words, we can perform the one-dimensional SSQW on a lattice which consists of only the even- (or the odd-) numbered lattice sites by performing two steps of the OQW with alternatively changing coin operators.  

\subsection{Decomposing two-dimensional SSQW}\label{Subsec:decomposition:2d}

A  decomposition similar to~\eqref{Eq:Zss-ZZ} can also be obtained for the two-dimensional SSQW propagator $Z_{\rm 2D}(\theta_1,\theta_2)$~\eqref{Eq:2DSS-U} in terms of two one-dimensional SSQWs performed on independent one-dimensional lattices. 
Using the  definitions~\eqref{Eq:2DSS-T1},~\eqref{Eq:2DSS-T2}, $S_3 = S_1S_2$, and the cyclic property of the quantum walk propagators we can simplify the propagator $Z_{\rm 2D}(\theta_1,\theta_2)$ as
    \begin{align}
      Z_{\rm 2D}(\theta_1,\theta_2) & = S_2\mathds{C}_{\theta_1}S_2 \mathds{C}_{\theta_2} S_1\mathds{C}_{\theta_1}S_1\\
                                    &= Z^{(2)}_{\rm ss}(0,\theta_1)Z^{(1)}_{\rm ss}(\theta_2,\theta_1),\label{Eq:Dec-2D}
    \end{align}
    where we have used the definition of the OQW propagator $Z(\theta)$~\eqref{Eq:Z} and the decomposition~\eqref{Eq:Zss-ZZ}.     Here the superscript $(i) \in \{(1),(2)\}$ denotes the DoF the operator is acting on. Equation~\eqref{Eq:Dec-2D} clearly shows that the two-dimensional SSQW on a triangular lattice can be decomposed into two one-dimensional SSQWs performed in series on two different DoFs.

  One of the advantages of the decompositions~\eqref{Eq:Zss-ZZ} and~\eqref{Eq:Dec-2D} is that now we can write the Hamiltonian $H_{\text{ss}}= \text{i}\ln[Z_{\text{ss}}(\theta_1,\theta_2)]$ and $H_{\text{2Dss}} = \text{i}\ln[Z_{\text{2D}}(\theta_1,\theta_2)]$ which govern the dynamics in one- and two-dimensional SSQWs as a function of the Hamiltonian $H_\theta = \text{i}\ln[Z(\theta)]$ of OQWs (see Appendix).  Hence, these decompositions offer an alternative way to express complicated  Hamiltonians in terms of simple well-understood one-dimensional quantum walk Hamiltonians.

    So far, we have shown a procedure to express one- and two-dimensional SSQWs using only OQWs on different DoFs or on different one-dimensional lattices. 
   In the following, we present optical implementation schemes to simulate an SSQW in the OAM and the time-bin space of light. We also propose methods to simulate Majorana modes, edge states, and the topologically protected bound states in these systems.

   \section{Optical implementation schemes for SSQW}\label{Sec:Implementation}
   Here we present optical implementation schemes to simulate the one- and the two-dimensional SSQWs. We also present a scheme to simulate the exotic phases such as Majorana modes and edge states. The schemes presented here are based on the decompositions constructed in the previous section, which make use of the one-dimensional OQW. Hence, our implementation scheme for a SSQW uses earlier schemes for OQWs~\cite{Schreiber2010,Schreiber2012,Goyal2013,Goyal2015a,Cardano2015,Cardano2016}.

   This section is organized as follows: In Sec.~\ref{Subsec:Implementation-1d} we detail the implementation of one-dimensional SSQW in OAM and in time-bins space. The implementation scheme for the two-dimensional SSQW is presented in Sec.~\ref{Subsec:Implementation-2d}. Section~\ref{Subsec:Simulation} contains the method to simulate Majorana modes and edge states in optical systems.

   \subsection{One-dimensional SSQW}\label{Subsec:Implementation-1d}
   The one-dimensional SSQW propagator $Z_{\text{ss}}$~\eqref{Eq:Zss} in the OAM space of light can be implemented by two propagators $Z(\theta_1)$ and $Z(\theta_2)$ in series~\eqref{Eq:Zss-ZZ}. To implement the OQW propagator $Z(\theta)$  in the OAM space we can use the scheme presented in~\cite{Goyal2013}.  In this scheme, the OAM states $\{\ket{\ell}, \ell\in \mathds{Z}\}$ represent the lattice sites and the right- ($\ket{R}$) and the left-handed ($\ket{L}$) circular polarization states of light  represent the two orthogonal states of the coin~\cite{Goyal2013}.
    The coin-flip operator $C_\theta$ is realized using the Simon-Mukunda polarization (SMP) gadget which consists of a combination of two half-wave plates and two quarter-wave plates mounted in series~\cite{Simon1989}. By rotating the wave plates one can realize  an arbitrary SU($2$)  rotation in the polarization states of light.

    The conditional propagator $S$~\eqref{Eq:S} in this scheme is realized by a combination of a half-wave plate  and an optical device called  $q$-plate.
The  $q$-plate is a linear optical device which couples the OAM of light with its polarization. It is a birefringent plate made of  a thin liquid crystal film sandwiched between glass substrates with a phase retardation $\delta$. Its inhomogeneous birefringence optical axis is distributed in space according to a singular pattern characterized by the topological charge $q$ which is the nematic-order defect exhibited in the center of the plate. Here $q$ can be an integer or half-integer number \cite{Marrucci2006,Slussarenko2011}. The action of the $q$-plate on the state of a light beam can be represented by the operator $Q_\delta^{(q)}$ as
\begin{align}
  Q_\delta^{(q)} =& \cos\delta \mathds{1}-\text{i}\sin\delta \big(F_{2q}\otimes\ket{L}\bra{R} + F^\dagger_{2q}\otimes\ket{R}\bra{L}\big),\label{Eq:QGen}
\end{align}
where $F_{2q} = \sum_\ell \ket{\ell+2q}\bra{\ell}$ represents the forward shift operator in the OAM of light. The phase retardation $\delta$ of the $q$-plate can be controlled by applying an external electric potential~\cite{Marrucci2006,Bliokh2011,Piccirillo2010} and can take any value between $0$ and $\pi$.

 The conditional  propagator $S$~\eqref{Eq:S} is realized  by setting  $q = 1/2$ and $\delta =\pi/2$~\cite{Marrucci2006,Bliokh2011,Piccirillo2010}. Thus, concatenating an SMP gadget, a $q$-plate, and a half-wave plate we can realize the OQW propagator $Z(\theta)$~\eqref{Eq:Z} which can perform a single step of quantum walk on the OAM of light. Placing two such setups in series in a ring interferometer, one can simulate the one-dimensional SSQW.

The one-dimensional SSQW in the time-bins space of light can be performed using the scheme presented in~\cite{Schreiber2010,Schreiber2012}.  In this scheme, the time-bins form the one-dimensional lattice and the polarization serves as the coin. The conditional propagator $S$~\eqref{Eq:S} is realized by splitting the incoming  light pulse (or single photon) into two spatial modes using a polarizing beam splitter and introducing different optical paths in the two spatial modes. The SMP gadget is used as the coin operator $C_\theta$. Thus, a single step of the OQW is performed by introducing different delays corresponding to the different orthogonal polarization states of light followed by an SMP gadget. Repeating this process twice with different SMP gadgets corresponding to the parameter $\theta_1$ and $\theta_2$ results in the one-dimensional SSQW. 

    Although, the relation~\eqref{Eq:Zss-ZZ} between the SSQWs and the OQW makes the implementation of the one-dimensional SSQW feasible in almost any quantum system, a much-simplified scheme can be achieved while realizing the quantum walk in the OAM space. 
 To see this, we recall the SSQW propagator $Z_{\text{ss}}(\theta_1,\theta_2)$ \eqref{Eq:Zss} which contains an effective shift operator $T_\theta$~\eqref{Eq:T-theta} and a coin-flip operator $C_\theta$. 
The operator $T_\theta$ can be realized using only a $q$-plate and wave plates by choosing the phase retardation $\delta =\pi/2 - \theta$ and $q = 1/2$ for the $q$-plate;
the operator $Q^{(q)}_\delta$~(see Appendix) reads
\begin{equation}
  Q^{(1/2)}_\delta = -\text{i}(\mathds{1}\otimes\sigma_x) \big[\cos\theta S + \text{i}\sin\theta(\mathds{1}\otimes \sigma_x) \big].
\end{equation}
Clearly, by redefining the coin-flip operator $C_{\theta_1}$ in the SSQW propagator $Z_{\rm ss}(\theta_1,\theta_2)$ in the following manner
    \begin{equation}
      \tilde{C}_{\theta_1} = e^{-\text{i}\pi\sigma_z/4}C_{\theta_1} \sigma_x e^{\text{i}\pi\sigma_z/4},
    \end{equation}
 which can be realized by an SMP device, we can realize the SSQW using a single $q$-plate and a single SMP gadget instead of using two of each as was done in the previous scheme.

    A similar scheme was  implemented by Cardano {\em et al.} in~\cite{Cardano2016} with a different coin operation. Although they also showed the topological order in their experiment, it was not clear if their experiment yielded the one-dimensional SSQW presented in~\cite{Kitagawa2010}. From our discussion, this question is now settled.

 Next, we present an implementation scheme to perform two-dimensional SSQWs on the triangular lattice.
\subsection{Two-dimensional SSQW}\label{Subsec:Implementation-2d}
\begin{figure}
  \includegraphics[width=0.45\textwidth]{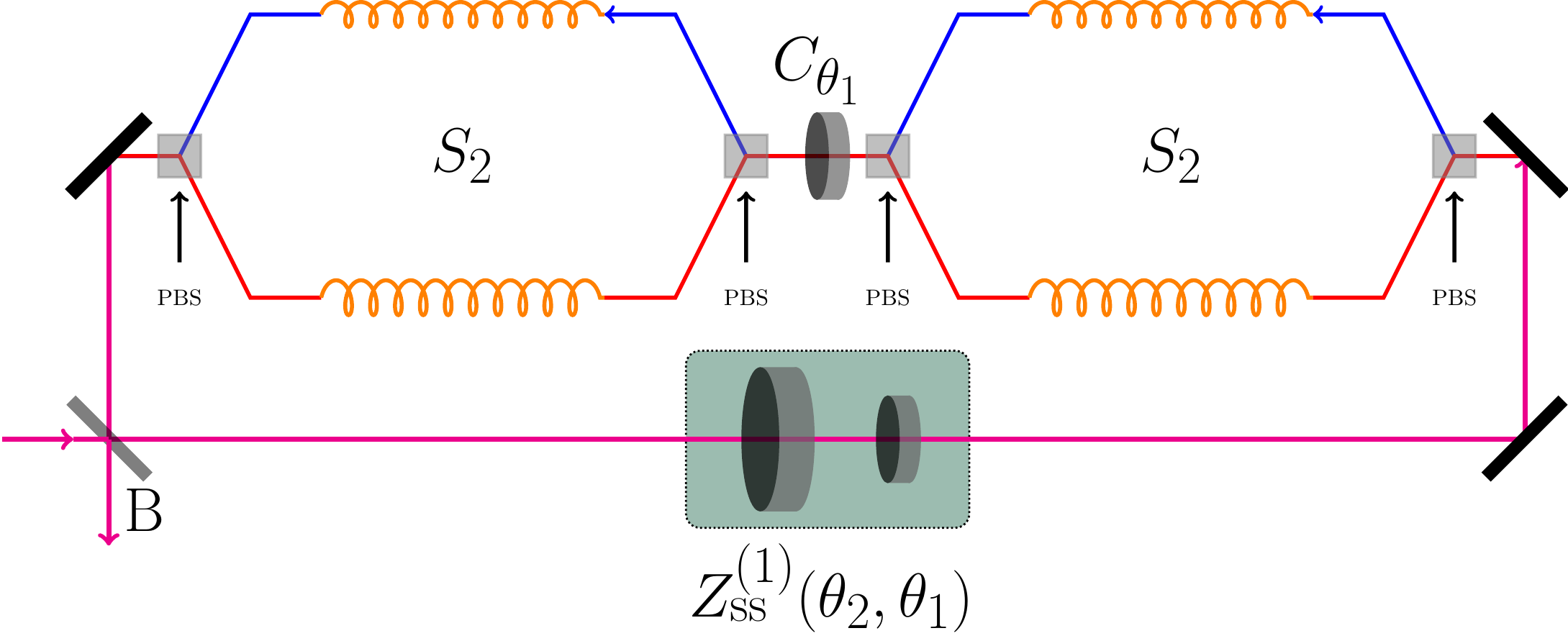}
  \caption{Optical implementation scheme for the two-dimensional SSQW based on Eq.~\eqref{Eq:Dec-2D}. Here the operator $Z^{(1)}_{\rm ss}(\theta_2,\theta_1)$ is performed over the OAM of light (lower part of the ring interferometer) and the operator $Z^{(2)}_{\rm ss}(0,\theta_1)$ is performed over the time-bins. In this scheme, the incoming light enters the setup through the beamsplitter B which has a very high reflectance $(r \approx 1)$ (bottom left corner). The light entering the setup is first transformed by a combination of an SMP gadget and a phase-retarded $q$-plate and then by two time-shift loops which are separated by an SMP gadget. The time-shift loops are realized by sorting the two orthogonal polarization components of light [using polarizing beamsplitters, (PBS)] in different spatial modes with unequal path lengths which can be achieved by multimode optical fibers.   One step of the two-dimensional SSQW is completed upon completion of a full circle in the setup. Upon returning to the beam splitter B a small fraction of the light will pass through the beam splitter and the rest will be reflected back in the setup. The transmitted part of the light can be used to perform real-time measurements on the OAM and the time-bins.
}
\label{Fig:Scheme2D}
\end{figure}

    In this scheme, we use both the OAM and the time-bins to perform two-dimensional SSQWs. We choose these DoFs because these are one of the most favored DoFs of light to perform quantum walks~\cite{Schreiber2010,Schreiber2012,Goyal2013,Goyal2015a,Cardano2015,Cardano2016}. A sketch of our implementation scheme is presented in Fig.~\ref{Fig:Scheme2D}.

    Here OAM is used as the first principal axis and time-bins are used as the second principal axis in the triangular lattice. Therefore, $S_1$ represents the conditional propagator on OAM and $S_2$ on time-bins. In order to realize the $Z_{\rm ss}^{(1)}(\theta_2,\theta_1)$ in Eq.~\eqref{Eq:Dec-2D} we use the $q$-plate and an SMP device as was done in the one-dimensional SSQW in OAM space. The $Z_{\rm ss}^{(2)}(0,\theta_1) = S_2C_{\theta_1}S_2$ operation in the quantum walk is performed over the time-bins space of light, as discussed earlier. The details of the scheme can be found in the caption of Fig.~\ref{Fig:Scheme2D}.

\subsection{Simulating Majorana modes and edge states}\label{Subsec:Simulation}
To simulate the topologically protected bound states in quantum walks we need to create a boundary with two distinct topological phases on either side by assigning  position-dependent values to the coin parameter $\theta$.  It is often hard to realize such coin operator. Here we propose a simple linear optical device, a generalized SMP gadget, which can be used to realize two different values of $\theta$ for different sections of the OAM lattice. 

 A light beam having the OAM proportional to the value $\ell\hbar$  has a ring-shaped intensity distribution with radius $r^\ell_{\rm max}$ of maximum intensity given by
\begin{equation}
  r^\ell_{\rm max} = w(z) \sqrt{\frac{|\ell|}{2}},
\end{equation}
i.e., the radius of the ring is proportional to the square root of the absolute value of $\ell$ of the OAM mode~\cite{Padgett1995}. Here $w(z)$ is the width of the laser beam at the position $z$. Therefore,  an SMP gadget with wave plates of radius $r_\ell = (r^\ell_{\rm max} + r^{\ell +1}_{\rm max})/2$ will cause rotations only for the polarization states of the OAM modes between $-\ell$ and $\ell$. The rest of the modes will remain unchanged.

\begin{figure}
  \includegraphics[width=0.3\textwidth]{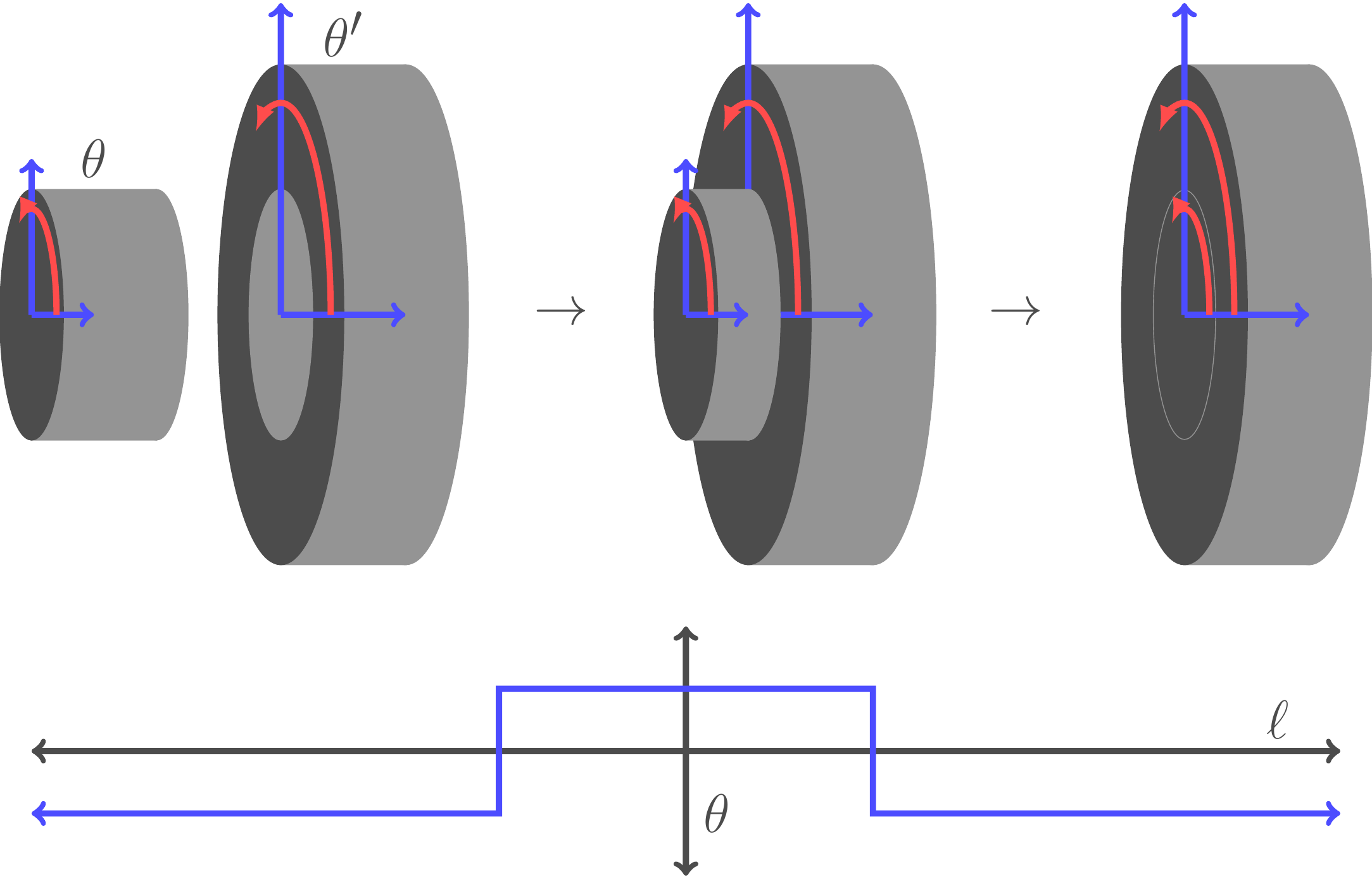}
  \caption{Generalized SMP gadget. A generalized SMP gadget is realized by coaxial placement of two SMP gadgets containing wave plates of different radii. The radius $r_\ell$ of the smaller SMP gadget is chosen such that it affects only the polarization states of light corresponding to the OAM modes between $-\ell$ and $\ell$. The rest of the polarization is transformed by the larger SMP gadget.   }
  \label{Fig:ModSMDevice}
\end{figure}

A generalized SMP gadget consists of two coaxial SMP gadgets, one with radius $r_\ell$ and the other, annular shaped, with inner radius $r_\ell$ and large outer radius (see Fig.~\ref{Fig:ModSMDevice}). Since the two SMP gadgets used here are independent of each other, they can be set to realize $C_\theta$ and $C_{\theta'}$ coin operations which result in different coin operations on different sections of the OAM lattice. 

Using the generalized SMP device, different values of the parameters $\theta$ and $\theta'$ can be chosen which correspond to distinct topological phases, thus making a boundary which supports bound states. 
In one-dimensional SSQWs, if initially the walker is localized at the boundary, it remains localized with large probability. These bound states correspond to Majorana modes. Since the two-dimensional SSQW can be decomposed into two one-dimensional SSQWs and we need the boundary only in one direction, we can use the same generalized SMP gadget to simulate the edge states in two-dimensional SSQWs. These bounded states can be observed as partially  localized states, i.e., localized  in the OAM space but spreading in the time-bin space. Both Majorana modes and topologically protected edge states are robust against environmental interactions.

Although the generalized SMP gadget provides a sharp transition in the value of the parameter $\theta$, it may not be sharp on the OAM lattice. This is because the radius of the intensity ring for a given OAM mode $\ell$ is not sharp. This may cause an aberration in the bound states if the parameter $\theta$ varies slightly across the boundary. However, if the two distinct topological phases require well-separated values of $\theta$ across the boundary, we will observe topologically protected bound states.

\section{Conclusion}\label{Sec:Conclusion}

To conclude, we have presented schemes to realize SSQWs on one- and two-dimensional lattices in optical systems. We have used the OAM and time-bin DoFs for our schemes. The key finding in this article which made these implementations feasible is the decomposition of SSQW in terms of OQWs. We have shown that a single step in a one-dimensional SSQW as defined in~\cite{Kitagawa2010} is nothing but two steps of the OQW with alternating coins. Similarly, a two-dimensional SSQW (on a triangular lattice) can be decomposed as two one-dimensional SSQWs performed on two independent DoFs in sequence. We have exploited this nature of SSQWs to simulate exotic topologically bound states.

We can also interpret the decomposition of SSQWs in terms  of the OQWs as follows: the Hamiltonian which governs the dynamics of a two-dimensional SSQW can be simulated by the Hamiltonians of a one-dimensional SSQW which in turn can be simulated by OQW Hamiltonians. This decomposition can be extended to realize more complicated quantum walks by incorporating multiple steps of the one-dimensional OQW with different coin parameters. Thus, our decomposition brings us a step closer to realizing a universal quantum simulator based purely on quantum walks.

\begin{acknowledgments}

SKG and CS acknowledge the support from NSERC. BCS thanks China's 1000 Talent Plan and the National Natural Science Foundation of China Grant No.~11675164, NSERC, and Alberta Innovates for financial support. WZ appreciates the financial support from China's 1000 Talent Plan and the National Natural Science Foundation of China Grant No.~11675164, the China Scholarship Council (Grant  No.~201406470022), and NSERC.
\end{acknowledgments}

\appendix

\section { Writing $H_{\text{ss}}$ and $H_{\text{2dss}}$ in terms of $H_\theta$.}
The Hamiltonian $H_\theta $ which governs the dynamics in the ordinary one-dimensional quantum walk can be calculated by
\begin{equation}
  H_\theta = \text{i}\log\big[Z(\theta)\big].
\end{equation}
Since the propagator $Z(\theta)$ is translation invariant, the Hamiltonian $H_\theta$ can be block diagonalized in momentum eigenbasis $\{\ket{k}\}$ which are
\begin{equation}
  \ket{k} = \sum_x \exp(-\text{i}kx)\ket{x},
\end{equation}
where $\{\ket{x}\}$ represents the position basis. In the momentum basis the Hamiltonian $H_\theta$ reads
\begin{equation}
  H_\theta = \bigoplus_{k\in[-\pi,\pi)}H(k),
\end{equation}
where
\begin{equation}
  H(k) = E(k)\bm{n}(k)\cdot \bm{\sigma}.
\end{equation}
The vector $\bm{n}(k) = [n_1(k),n_2(k),n_3(k)]$ is a three-dimensional real vector. The explicit form of the vector $\bm{n}(k)$ and the energy $E(k)$ is given by
\begin{align}
  n_1(k) &= \frac{\sin\theta\sin k}{\sin[E(k)]},\\
  n_2(k) &= \frac{\sin\theta\cos k}{\sin[E(k)]},\\
  n_3(k) &= -\frac{\cos\theta\sin k}{\sin[E(k)]},\\
  E(k) &= \cos^{-1}(\cos\theta\cos k).
\end{align}

The relation between the Hamiltonians $H_{\text{ss}}$ and $H_\theta$ can be derived using the decomposition~\eqref{Eq:Zss-ZZ} and the  Baker-Campbell-Hausdorff formula~\cite{Hornmatrix}
\onecolumngrid
    \begin{align}
      H_\text{ss}(k) =& H_{\theta_1}(k) + H_{\theta_2}(k) - \frac{\text{i}}{2}[H_{\theta_1}(k),H_{\theta_2}(k)]- \frac{1}{12}\{[H_{\theta_1}(k),[H_{\theta_1}(k),H_{\theta_2}(k)]] + [H_{\theta_2}(k),[H_{\theta_2}(k),H_{\theta_1}(k)]]\} \cdots,
    \end{align}

    whereas $H_{\text{ss}} = \text{i} \ln[Z_{\text{ss}}(\theta_1,\theta_2)]$ yields
    \begin{align}
H_{\text{ss}}(k)      =& \frac{E_{\text{ss}}(k)}{\sin[E_{\text{ss}}(k)]}\left[ \cos E_1(k)\sin E_2(k) \bm{n}_2(k) + \cos E_2(k)\sin E_1(k) \bm{n}_1(k) + \sin E_1(k) \sin E_2(k) \bm{n}_1(k)\times \bm{n}_2(k)\right].\bm{\sigma}\\
      \equiv& E_{\text{ss}}(k)\bm{N}(k)\cdot \bm{\sigma},\\ 
      E_{\text{ss}}(k) =& \cos^{-1}\left[\cos E_1(k) \cos E_2(k) - \sin E_1(k) \sin E_2(k) \bm{n}_1(k)\cdot\bm{n}_2(k)\right],\\
      E_i(k) =& \cos^{-1}(\cos\theta_i\cos k).
    \end{align}
    Here $E_i(k)\bm{n}_i\cdot\bm{\sigma} = H(\theta_i)$.
    
    Similarly, the Hamiltonian $H_{\text{2Dss}}$ can be written in terms of $H_{\text{ss}}$ as
    \begin{align}
      H_\text{2dss}(k_x,k_y) =& H_\text{ss}(k_x) + H_\text{ss}(k_y) - \frac{\text{i}}{2}[H_{\text{ss}}(k_x),H_{\text{ss}}(k_y)] \nonumber\\
      &\qquad \qquad- \frac{1}{12}\{[H_{\text{ss}}(k_x),[H_{\text{ss}}(k_x),H_{\text{ss}}(k_y)]] + [H_{\text{ss}}(k_y),[H_{\text{ss}}(k_y),H_{\text{ss}}(k_x)]]\} \cdots\\
      =& \frac{E_{\text{2Dss}}(k_x,k_y)}{\sin(E_{\text{2Dss}}(k_x,k_y))}\left( \cos E_{\text{ss}}(k_x)\sin E_{\text{ss}}(k_y) \bm{N}(k_y) + \cos E_{\text{ss}}(k_y)\sin E_{\text{ss}}(k_x) \bm{N}(k_x)\right.\nonumber\\
      &\qquad \qquad \qquad\left.+ \sin E_{\text{ss}}(k_x) \sin E_{\text{ss}}(k_y) \bm{N}(k_x)\times \bm{N}(k_y)\right).\bm{\sigma},\\
      E_{\text{2Dss}}(k_x,k_y) =& \cos^{-1}\left[\cos E_{\text{ss}}(k_x) \cos E_{\text{ss}}(k_y) - \sin E_{\text{ss}}(k_x) \sin E_{\text{ss}}(k_y) \bm{N}(k_x)\cdot\bm{N}(k_y)\right].
    \end{align}
\twocolumngrid
    

\begin{thebibliography}{46}%
\makeatletter
\providecommand \@ifxundefined [1]{%
 \@ifx{#1\undefined}
}%
\providecommand \@ifnum [1]{%
 \ifnum #1\expandafter \@firstoftwo
 \else \expandafter \@secondoftwo
 \fi
}%
\providecommand \@ifx [1]{%
 \ifx #1\expandafter \@firstoftwo
 \else \expandafter \@secondoftwo
 \fi
}%
\providecommand \natexlab [1]{#1}%
\providecommand \enquote  [1]{``#1''}%
\providecommand \bibnamefont  [1]{#1}%
\providecommand \bibfnamefont [1]{#1}%
\providecommand \citenamefont [1]{#1}%
\providecommand \href@noop [0]{\@secondoftwo}%
\providecommand \href [0]{\begingroup \@sanitize@url \@href}%
\providecommand \@href[1]{\@@startlink{#1}\@@href}%
\providecommand \@@href[1]{\endgroup#1\@@endlink}%
\providecommand \@sanitize@url [0]{\catcode `\\12\catcode `\$12\catcode
  `\&12\catcode `\#12\catcode `\^12\catcode `\_12\catcode `\%12\relax}%
\providecommand \@@startlink[1]{}%
\providecommand \@@endlink[0]{}%
\providecommand \url  [0]{\begingroup\@sanitize@url \@url }%
\providecommand \@url [1]{\endgroup\@href {#1}{\urlprefix }}%
\providecommand \urlprefix  [0]{URL }%
\providecommand \Eprint [0]{\href }%
\providecommand \doibase [0]{http://dx.doi.org/}%
\providecommand \selectlanguage [0]{\@gobble}%
\providecommand \bibinfo  [0]{\@secondoftwo}%
\providecommand \bibfield  [0]{\@secondoftwo}%
\providecommand \translation [1]{[#1]}%
\providecommand \BibitemOpen [0]{}%
\providecommand \bibitemStop [0]{}%
\providecommand \bibitemNoStop [0]{.\EOS\space}%
\providecommand \EOS [0]{\spacefactor3000\relax}%
\providecommand \BibitemShut  [1]{\csname bibitem#1\endcsname}%
\let\auto@bib@innerbib\@empty
\bibitem [{\citenamefont {{von Klitzing}}\ \emph {et~al.}(1980)\citenamefont
  {{von Klitzing}}, \citenamefont {Dorda},\ and\ \citenamefont
  {Pepper}}]{Klitzing1980}%
  \BibitemOpen
  \bibfield  {author} {\bibinfo {author} {\bibfnamefont {K.}~\bibnamefont {{von
  Klitzing}}}, \bibinfo {author} {\bibfnamefont {G.}~\bibnamefont {Dorda}}, \
  and\ \bibinfo {author} {\bibfnamefont {M.}~\bibnamefont {Pepper}},\ }\href
  {\doibase 10.1103/physrevlett.45.494} {\bibfield  {journal} {\bibinfo
  {journal} {Phys. Rev. Lett.}\ }\textbf {\bibinfo {volume} {45}},\ \bibinfo
  {pages} {494} (\bibinfo {year} {1980})}\BibitemShut {NoStop}%
\bibitem [{\citenamefont {Thouless}\ \emph {et~al.}(1982)\citenamefont
  {Thouless}, \citenamefont {Kohmoto}, \citenamefont {Nightingale},\ and\
  \citenamefont {den Nijs}}]{Thouless1982}%
  \BibitemOpen
  \bibfield  {author} {\bibinfo {author} {\bibfnamefont {D.~J.}\ \bibnamefont
  {Thouless}}, \bibinfo {author} {\bibfnamefont {M.}~\bibnamefont {Kohmoto}},
  \bibinfo {author} {\bibfnamefont {M.~P.}\ \bibnamefont {Nightingale}}, \ and\
  \bibinfo {author} {\bibfnamefont {M.}~\bibnamefont {den Nijs}},\ }\href
  {\doibase 10.1103/physrevlett.49.405} {\bibfield  {journal} {\bibinfo
  {journal} {Phys. Rev. Lett.}\ }\textbf {\bibinfo {volume} {49}},\ \bibinfo
  {pages} {405} (\bibinfo {year} {1982})}\BibitemShut {NoStop}%
\bibitem [{\citenamefont {Laughlin}(1983)}]{Laughlin1983}%
  \BibitemOpen
  \bibfield  {author} {\bibinfo {author} {\bibfnamefont {R.~B.}\ \bibnamefont
  {Laughlin}},\ }\href {\doibase 10.1103/physrevlett.50.1395} {\bibfield
  {journal} {\bibinfo  {journal} {Phys. Rev. Lett.}\ }\textbf {\bibinfo
  {volume} {50}},\ \bibinfo {pages} {1395} (\bibinfo {year}
  {1983})}\BibitemShut {NoStop}%
\bibitem [{\citenamefont {Wen}(1995)}]{Wen1995}%
  \BibitemOpen
  \bibfield  {author} {\bibinfo {author} {\bibfnamefont {X.-G.}\ \bibnamefont
  {Wen}},\ }\href {\doibase 10.1080/00018739500101566} {\bibfield  {journal}
  {\bibinfo  {journal} {Adv. Phys.}\ }\textbf {\bibinfo {volume} {44}},\
  \bibinfo {pages} {405} (\bibinfo {year} {1995})}\BibitemShut {NoStop}%
\bibitem [{\citenamefont {Stormer}\ \emph {et~al.}(1999)\citenamefont
  {Stormer}, \citenamefont {Tsui},\ and\ \citenamefont
  {Gossard}}]{Stormer1999}%
  \BibitemOpen
  \bibfield  {author} {\bibinfo {author} {\bibfnamefont {H.~L.}\ \bibnamefont
  {Stormer}}, \bibinfo {author} {\bibfnamefont {D.~C.}\ \bibnamefont {Tsui}}, \
  and\ \bibinfo {author} {\bibfnamefont {A.~C.}\ \bibnamefont {Gossard}},\
  }\href {\doibase 10.1103/revmodphys.71.s298} {\bibfield  {journal} {\bibinfo
  {journal} {Rev. Mod. Phys.}\ }\textbf {\bibinfo {volume} {71}},\ \bibinfo
  {pages} {S298} (\bibinfo {year} {1999})}\BibitemShut {NoStop}%
\bibitem [{\citenamefont {Ryu}\ and\ \citenamefont {Hatsugai}(2002)}]{Ryu2002}%
  \BibitemOpen
  \bibfield  {author} {\bibinfo {author} {\bibfnamefont {S.}~\bibnamefont
  {Ryu}}\ and\ \bibinfo {author} {\bibfnamefont {Y.}~\bibnamefont {Hatsugai}},\
  }\href {\doibase 10.1103/physrevlett.89.077002} {\bibfield  {journal}
  {\bibinfo  {journal} {Phys. Rev. Lett.}\ }\textbf {\bibinfo {volume} {89}},\
  \bibinfo {pages} {077002} (\bibinfo {year} {2002})}\BibitemShut {NoStop}%
\bibitem [{\citenamefont {Kane}\ and\ \citenamefont
  {Mele}(2005{\natexlab{a}})}]{Kane2005}%
  \BibitemOpen
  \bibfield  {author} {\bibinfo {author} {\bibfnamefont {C.~L.}\ \bibnamefont
  {Kane}}\ and\ \bibinfo {author} {\bibfnamefont {E.~J.}\ \bibnamefont
  {Mele}},\ }\href {\doibase 10.1103/PhysRevLett.95.146802} {\bibfield
  {journal} {\bibinfo  {journal} {Phys. Rev. Lett.}\ }\textbf {\bibinfo
  {volume} {95}},\ \bibinfo {pages} {146802} (\bibinfo {year}
  {2005}{\natexlab{a}})}\BibitemShut {NoStop}%
\bibitem [{\citenamefont {Kane}\ and\ \citenamefont
  {Mele}(2005{\natexlab{b}})}]{Kane2005a}%
  \BibitemOpen
  \bibfield  {author} {\bibinfo {author} {\bibfnamefont {C.~L.}\ \bibnamefont
  {Kane}}\ and\ \bibinfo {author} {\bibfnamefont {E.~J.}\ \bibnamefont
  {Mele}},\ }\href {\doibase 10.1103/physrevlett.95.226801} {\bibfield
  {journal} {\bibinfo  {journal} {Phys. Rev. Lett.}\ }\textbf {\bibinfo
  {volume} {95}},\ \bibinfo {pages} {226801} (\bibinfo {year}
  {2005}{\natexlab{b}})}\BibitemShut {NoStop}%
\bibitem [{\citenamefont {Hsieh}\ \emph {et~al.}(2008)\citenamefont {Hsieh},
  \citenamefont {Qian}, \citenamefont {Wray}, \citenamefont {Xia},
  \citenamefont {Hor}, \citenamefont {Cava},\ and\ \citenamefont
  {Hasan}}]{Hsieh2008}%
  \BibitemOpen
  \bibfield  {author} {\bibinfo {author} {\bibfnamefont {D.}~\bibnamefont
  {Hsieh}}, \bibinfo {author} {\bibfnamefont {D.}~\bibnamefont {Qian}},
  \bibinfo {author} {\bibfnamefont {L.}~\bibnamefont {Wray}}, \bibinfo {author}
  {\bibfnamefont {Y.}~\bibnamefont {Xia}}, \bibinfo {author} {\bibfnamefont
  {Y.~S.}\ \bibnamefont {Hor}}, \bibinfo {author} {\bibfnamefont {R.~J.}\
  \bibnamefont {Cava}}, \ and\ \bibinfo {author} {\bibfnamefont {M.~Z.}\
  \bibnamefont {Hasan}},\ }\href {\doibase 10.1038/nature06843} {\bibfield
  {journal} {\bibinfo  {journal} {Nature}\ }\textbf {\bibinfo {volume} {452}},\
  \bibinfo {pages} {970} (\bibinfo {year} {2008})}\BibitemShut {NoStop}%
\bibitem [{\citenamefont {Hasan}\ and\ \citenamefont {Kane}(2010)}]{Hasan2010}%
  \BibitemOpen
  \bibfield  {author} {\bibinfo {author} {\bibfnamefont {M.~Z.}\ \bibnamefont
  {Hasan}}\ and\ \bibinfo {author} {\bibfnamefont {C.~L.}\ \bibnamefont
  {Kane}},\ }\href {\doibase 10.1103/revmodphys.82.3045} {\bibfield  {journal}
  {\bibinfo  {journal} {Rev. Mod. Phys.}\ }\textbf {\bibinfo {volume} {82}},\
  \bibinfo {pages} {3045} (\bibinfo {year} {2010})}\BibitemShut {NoStop}%
\bibitem [{\citenamefont {Bernevig}\ \emph {et~al.}(2006)\citenamefont
  {Bernevig}, \citenamefont {Hughes},\ and\ \citenamefont
  {Zhang}}]{Bernevig2006}%
  \BibitemOpen
  \bibfield  {author} {\bibinfo {author} {\bibfnamefont {B.~A.}\ \bibnamefont
  {Bernevig}}, \bibinfo {author} {\bibfnamefont {T.~L.}\ \bibnamefont
  {Hughes}}, \ and\ \bibinfo {author} {\bibfnamefont {S.-C.}\ \bibnamefont
  {Zhang}},\ }\href {\doibase 10.1126/science.1133734} {\bibfield  {journal}
  {\bibinfo  {journal} {Science}\ }\textbf {\bibinfo {volume} {314}},\ \bibinfo
  {pages} {1757} (\bibinfo {year} {2006})}\BibitemShut {NoStop}%
\bibitem [{\citenamefont {Nayak}\ \emph {et~al.}(2008)\citenamefont {Nayak},
  \citenamefont {Simon}, \citenamefont {Stern}, \citenamefont {Freedman},\ and\
  \citenamefont {Das~Sarma}}]{Nayak2008}%
  \BibitemOpen
  \bibfield  {author} {\bibinfo {author} {\bibfnamefont {C.}~\bibnamefont
  {Nayak}}, \bibinfo {author} {\bibfnamefont {S.~H.}\ \bibnamefont {Simon}},
  \bibinfo {author} {\bibfnamefont {A.}~\bibnamefont {Stern}}, \bibinfo
  {author} {\bibfnamefont {M.}~\bibnamefont {Freedman}}, \ and\ \bibinfo
  {author} {\bibfnamefont {S.}~\bibnamefont {Das~Sarma}},\ }\href {\doibase
  10.1103/RevModPhys.80.1083} {\bibfield  {journal} {\bibinfo  {journal} {Rev.
  Mod. Phys.}\ }\textbf {\bibinfo {volume} {80}},\ \bibinfo {pages} {1083}
  (\bibinfo {year} {2008})}\BibitemShut {NoStop}%
\bibitem [{\citenamefont {Kitaev}(2003)}]{Kitaev2003}%
  \BibitemOpen
  \bibfield  {author} {\bibinfo {author} {\bibfnamefont {A.}~\bibnamefont
  {Kitaev}},\ }\href {\doibase 10.1016/s0003-4916(02)00018-0} {\bibfield
  {journal} {\bibinfo  {journal} {Ann. Phys.}\ }\textbf {\bibinfo {volume}
  {303}},\ \bibinfo {pages} {2} (\bibinfo {year} {2003})}\BibitemShut {NoStop}%
\bibitem [{\citenamefont {Kitaev}(2006)}]{Kitaev2006}%
  \BibitemOpen
  \bibfield  {author} {\bibinfo {author} {\bibfnamefont {A.}~\bibnamefont
  {Kitaev}},\ }\href {\doibase 10.1016/j.aop.2005.10.005} {\bibfield  {journal}
  {\bibinfo  {journal} {Ann. Phys.}\ }\textbf {\bibinfo {volume} {321}},\
  \bibinfo {pages} {2} (\bibinfo {year} {2006})}\BibitemShut {NoStop}%
\bibitem [{\citenamefont {Kitagawa}\ \emph {et~al.}(2010)\citenamefont
  {Kitagawa}, \citenamefont {Rudner}, \citenamefont {Berg},\ and\ \citenamefont
  {Demler}}]{Kitagawa2010}%
  \BibitemOpen
  \bibfield  {author} {\bibinfo {author} {\bibfnamefont {T.}~\bibnamefont
  {Kitagawa}}, \bibinfo {author} {\bibfnamefont {M.~S.}\ \bibnamefont
  {Rudner}}, \bibinfo {author} {\bibfnamefont {E.}~\bibnamefont {Berg}}, \ and\
  \bibinfo {author} {\bibfnamefont {E.}~\bibnamefont {Demler}},\ }\href
  {\doibase 10.1103/physreva.82.033429} {\bibfield  {journal} {\bibinfo
  {journal} {Phys. Rev. A}\ }\textbf {\bibinfo {volume} {82}},\ \bibinfo
  {pages} {033429} (\bibinfo {year} {2010})}\BibitemShut {NoStop}%
\bibitem [{\citenamefont {Obuse}\ and\ \citenamefont
  {Kawakami}(2011)}]{Obuse2011}%
  \BibitemOpen
  \bibfield  {author} {\bibinfo {author} {\bibfnamefont {H.}~\bibnamefont
  {Obuse}}\ and\ \bibinfo {author} {\bibfnamefont {N.}~\bibnamefont
  {Kawakami}},\ }\href {\doibase 10.1103/physrevb.84.195139} {\bibfield
  {journal} {\bibinfo  {journal} {Phys. Rev. B}\ }\textbf {\bibinfo {volume}
  {84}},\ \bibinfo {pages} {195139} (\bibinfo {year} {2011})}\BibitemShut
  {NoStop}%
\bibitem [{\citenamefont {Asb{\'o}th}(2012)}]{Asboth2012}%
  \BibitemOpen
  \bibfield  {author} {\bibinfo {author} {\bibfnamefont {J.~K.}\ \bibnamefont
  {Asb{\'o}th}},\ }\href {\doibase 10.1103/physrevb.86.195414} {\bibfield
  {journal} {\bibinfo  {journal} {Phys. Rev. B}\ }\textbf {\bibinfo {volume}
  {86}},\ \bibinfo {pages} {195414} (\bibinfo {year} {2012})}\BibitemShut
  {NoStop}%
\bibitem [{\citenamefont {Rakovszky}\ and\ \citenamefont
  {Asboth}(2015)}]{Rakovszky2015}%
  \BibitemOpen
  \bibfield  {author} {\bibinfo {author} {\bibfnamefont {T.}~\bibnamefont
  {Rakovszky}}\ and\ \bibinfo {author} {\bibfnamefont {J.~K.}\ \bibnamefont
  {Asboth}},\ }\href {\doibase 10.1103/physreva.92.052311} {\bibfield
  {journal} {\bibinfo  {journal} {Phys. Rev. A}\ }\textbf {\bibinfo {volume}
  {92}},\ \bibinfo {pages} {052311} (\bibinfo {year} {2015})}\BibitemShut
  {NoStop}%
\bibitem [{\citenamefont {Cedzich}\ \emph {et~al.}(2016)\citenamefont
  {Cedzich}, \citenamefont {Gr{\"u}nbaum}, \citenamefont {Stahl}, \citenamefont
  {Vel{\'a}zquez}, \citenamefont {Werner},\ and\ \citenamefont
  {Werner}}]{Cedzich2015}%
  \BibitemOpen
  \bibfield  {author} {\bibinfo {author} {\bibfnamefont {C.}~\bibnamefont
  {Cedzich}}, \bibinfo {author} {\bibfnamefont {F.}~\bibnamefont
  {Gr{\"u}nbaum}}, \bibinfo {author} {\bibfnamefont {C.}~\bibnamefont {Stahl}},
  \bibinfo {author} {\bibfnamefont {L.}~\bibnamefont {Vel{\'a}zquez}}, \bibinfo
  {author} {\bibfnamefont {A.}~\bibnamefont {Werner}}, \ and\ \bibinfo {author}
  {\bibfnamefont {R.}~\bibnamefont {Werner}},\ }\href
  {http://stacks.iop.org/1751-8121/49/i=21/a=21LT01} {\bibfield  {journal}
  {\bibinfo  {journal} {J. Phys. A: Math. Theor.}\ }\textbf {\bibinfo {volume}
  {49}},\ \bibinfo {pages} {21LT01} (\bibinfo {year} {2016})}\BibitemShut
  {NoStop}%
\bibitem [{\citenamefont {Obuse}\ \emph {et~al.}(2015)\citenamefont {Obuse},
  \citenamefont {Asb\'oth}, \citenamefont {Nishimura},\ and\ \citenamefont
  {Kawakami}}]{Obuse2015}%
  \BibitemOpen
  \bibfield  {author} {\bibinfo {author} {\bibfnamefont {H.}~\bibnamefont
  {Obuse}}, \bibinfo {author} {\bibfnamefont {J.~K.}\ \bibnamefont {Asb\'oth}},
  \bibinfo {author} {\bibfnamefont {Y.}~\bibnamefont {Nishimura}}, \ and\
  \bibinfo {author} {\bibfnamefont {N.}~\bibnamefont {Kawakami}},\ }\href
  {\doibase 10.1103/physrevb.92.045424} {\bibfield  {journal} {\bibinfo
  {journal} {Phys. Rev. B}\ }\textbf {\bibinfo {volume} {92}},\ \bibinfo
  {pages} {045424} (\bibinfo {year} {2015})}\BibitemShut {NoStop}%
\bibitem [{\citenamefont {Groh}\ \emph {et~al.}(2016)\citenamefont {Groh},
  \citenamefont {Brakhane}, \citenamefont {Alt}, \citenamefont {Meschede},
  \citenamefont {Asb{\'o}th},\ and\ \citenamefont {Alberti}}]{Groh2016}%
  \BibitemOpen
  \bibfield  {author} {\bibinfo {author} {\bibfnamefont {T.}~\bibnamefont
  {Groh}}, \bibinfo {author} {\bibfnamefont {S.}~\bibnamefont {Brakhane}},
  \bibinfo {author} {\bibfnamefont {W.}~\bibnamefont {Alt}}, \bibinfo {author}
  {\bibfnamefont {D.}~\bibnamefont {Meschede}}, \bibinfo {author}
  {\bibfnamefont {J.}~\bibnamefont {Asb{\'o}th}}, \ and\ \bibinfo {author}
  {\bibfnamefont {A.}~\bibnamefont {Alberti}},\ }\href {\doibase
  10.1103/physreva.94.013620} {\bibfield  {journal} {\bibinfo  {journal} {Phys.
  Rev. A}\ }\textbf {\bibinfo {volume} {94}},\ \bibinfo {pages} {013620}
  (\bibinfo {year} {2016})}\BibitemShut {NoStop}%
\bibitem [{\citenamefont {Lam}\ \emph {et~al.}(2016)\citenamefont {Lam},
  \citenamefont {Yu},\ and\ \citenamefont {Szeto}}]{Lam2016}%
  \BibitemOpen
  \bibfield  {author} {\bibinfo {author} {\bibfnamefont {H.~T.}\ \bibnamefont
  {Lam}}, \bibinfo {author} {\bibfnamefont {Y.}~\bibnamefont {Yu}}, \ and\
  \bibinfo {author} {\bibfnamefont {K.~Y.}\ \bibnamefont {Szeto}},\ }\href
  {\doibase 10.1103/physreva.93.052319} {\bibfield  {journal} {\bibinfo
  {journal} {Phys. Rev. A}\ }\textbf {\bibinfo {volume} {93}},\ \bibinfo
  {pages} {052319} (\bibinfo {year} {2016})}\BibitemShut {NoStop}%
\bibitem [{\citenamefont {Wilczek}\ and\ \citenamefont
  {Esposito}()}]{Wilczek2014}%
  \BibitemOpen
  \bibfield  {author} {\bibinfo {author} {\bibfnamefont {F.}~\bibnamefont
  {Wilczek}}\ and\ \bibinfo {author} {\bibfnamefont {S.}~\bibnamefont
  {Esposito}},\ }\href {\doibase 10.1017/cbo9781107358362.014} {\emph {\bibinfo
  {title} {Majorana and condensed matter physics}}}\ (\bibinfo  {publisher}
  {Cambridge University Press, New York, 2014})\ pp.\ \bibinfo {pages}
  {279--302}\BibitemShut {NoStop}%
\bibitem [{\citenamefont {Tamm}(1932)}]{Tamm1932}%
  \BibitemOpen
  \bibfield  {author} {\bibinfo {author} {\bibfnamefont {I.}~\bibnamefont
  {Tamm}},\ }\href@noop {} {\bibfield  {journal} {\bibinfo  {journal} {Phys. Z.
  Soviet Union}\ }\textbf {\bibinfo {volume} {1}},\ \bibinfo {pages} {732}
  (\bibinfo {year} {1932})}\BibitemShut {NoStop}%
\bibitem [{\citenamefont {Shockley}(1939)}]{Shockley1939}%
  \BibitemOpen
  \bibfield  {author} {\bibinfo {author} {\bibfnamefont {W.}~\bibnamefont
  {Shockley}},\ }\href {\doibase 10.1103/physrev.56.317} {\bibfield  {journal}
  {\bibinfo  {journal} {Phys. Rev.}\ }\textbf {\bibinfo {volume} {56}},\
  \bibinfo {pages} {317} (\bibinfo {year} {1939})}\BibitemShut {NoStop}%
\bibitem [{\citenamefont {Balachandran}\ \emph {et~al.}(1997)\citenamefont
  {Balachandran}, \citenamefont {Momen},\ and\ \citenamefont
  {Chandar}}]{Balachandran1997}%
  \BibitemOpen
  \bibfield  {author} {\bibinfo {author} {\bibfnamefont {A.~P.}\ \bibnamefont
  {Balachandran}}, \bibinfo {author} {\bibfnamefont {A.}~\bibnamefont {Momen}},
  \ and\ \bibinfo {author} {\bibfnamefont {L.}~\bibnamefont {Chandar}},\ }\href
  {\doibase 10.1142/s0217751x97000578} {\bibfield  {journal} {\bibinfo
  {journal} {Int. J. Mod. Phys. A}\ }\textbf {\bibinfo {volume} {12}},\
  \bibinfo {pages} {625} (\bibinfo {year} {1997})}\BibitemShut {NoStop}%
\bibitem [{\citenamefont {Corichi}(1999)}]{Corichi1999}%
  \BibitemOpen
  \bibfield  {author} {\bibinfo {author} {\bibfnamefont {A.}~\bibnamefont
  {Corichi}},\ }\href@noop {} {\bibfield  {journal} {\bibinfo  {journal} {Gen.
  Relativ. Gravit.}\ }\textbf {\bibinfo {volume} {31}},\ \bibinfo {pages} {615}
  (\bibinfo {year} {1999})}\BibitemShut {NoStop}%
\bibitem [{\citenamefont {Kitagawa}\ \emph {et~al.}(2012)\citenamefont
  {Kitagawa}, \citenamefont {Broome}, \citenamefont {Fedrizzi}, \citenamefont
  {Rudner}, \citenamefont {Berg}, \citenamefont {Kassal}, \citenamefont
  {Aspuru-Guzik}, \citenamefont {Demler},\ and\ \citenamefont
  {White}}]{Kitagawa2012}%
  \BibitemOpen
  \bibfield  {author} {\bibinfo {author} {\bibfnamefont {T.}~\bibnamefont
  {Kitagawa}}, \bibinfo {author} {\bibfnamefont {M.~A.}\ \bibnamefont
  {Broome}}, \bibinfo {author} {\bibfnamefont {A.}~\bibnamefont {Fedrizzi}},
  \bibinfo {author} {\bibfnamefont {M.~S.}\ \bibnamefont {Rudner}}, \bibinfo
  {author} {\bibfnamefont {E.}~\bibnamefont {Berg}}, \bibinfo {author}
  {\bibfnamefont {I.}~\bibnamefont {Kassal}}, \bibinfo {author} {\bibfnamefont
  {A.}~\bibnamefont {Aspuru-Guzik}}, \bibinfo {author} {\bibfnamefont
  {E.}~\bibnamefont {Demler}}, \ and\ \bibinfo {author} {\bibfnamefont {A.~G.}\
  \bibnamefont {White}},\ }\href {\doibase 10.1038/ncomms1872} {\bibfield
  {journal} {\bibinfo  {journal} {Nat. Commun.}\ }\textbf {\bibinfo {volume}
  {3}},\ \bibinfo {pages} {882} (\bibinfo {year} {2012})}\BibitemShut {NoStop}%
\bibitem [{\citenamefont {Di~Franco}\ \emph
  {et~al.}(2011{\natexlab{a}})\citenamefont {Di~Franco}, \citenamefont
  {Mc~Gettrick},\ and\ \citenamefont {Busch}}]{DiFranco2011}%
  \BibitemOpen
  \bibfield  {author} {\bibinfo {author} {\bibfnamefont {C.}~\bibnamefont
  {Di~Franco}}, \bibinfo {author} {\bibfnamefont {M.}~\bibnamefont
  {Mc~Gettrick}}, \ and\ \bibinfo {author} {\bibfnamefont {T.}~\bibnamefont
  {Busch}},\ }\href {\doibase 10.1103/physrevlett.106.080502} {\bibfield
  {journal} {\bibinfo  {journal} {Phys.~Rev.~Lett.}\ }\textbf {\bibinfo
  {volume} {106}},\ \bibinfo {pages} {080502} (\bibinfo {year}
  {2011}{\natexlab{a}})}\BibitemShut {NoStop}%
\bibitem [{\citenamefont {Di~Franco}\ \emph
  {et~al.}(2011{\natexlab{b}})\citenamefont {Di~Franco}, \citenamefont
  {Mc~Gettrick}, \citenamefont {Machida},\ and\ \citenamefont
  {Busch}}]{DiFranco2011a}%
  \BibitemOpen
  \bibfield  {author} {\bibinfo {author} {\bibfnamefont {C.}~\bibnamefont
  {Di~Franco}}, \bibinfo {author} {\bibfnamefont {M.}~\bibnamefont
  {Mc~Gettrick}}, \bibinfo {author} {\bibfnamefont {T.}~\bibnamefont
  {Machida}}, \ and\ \bibinfo {author} {\bibfnamefont {T.}~\bibnamefont
  {Busch}},\ }\href {\doibase 10.1103/physreva.84.042337} {\bibfield  {journal}
  {\bibinfo  {journal} {Phys.~Rev.~A}\ }\textbf {\bibinfo {volume} {84}},\
  \bibinfo {pages} {042337} (\bibinfo {year} {2011}{\natexlab{b}})}\BibitemShut
  {NoStop}%
\bibitem [{\citenamefont {Rold\'an}\ \emph {et~al.}(2013)\citenamefont
  {Rold\'an}, \citenamefont {Di~Franco}, \citenamefont {Silva},\ and\
  \citenamefont {de~Valc\'arcel}}]{Roldan2013}%
  \BibitemOpen
  \bibfield  {author} {\bibinfo {author} {\bibfnamefont {E.}~\bibnamefont
  {Rold\'an}}, \bibinfo {author} {\bibfnamefont {C.}~\bibnamefont {Di~Franco}},
  \bibinfo {author} {\bibfnamefont {F.}~\bibnamefont {Silva}}, \ and\ \bibinfo
  {author} {\bibfnamefont {G.~J.}\ \bibnamefont {de~Valc\'arcel}},\ }\href
  {\doibase 10.1103/physreva.87.022336} {\bibfield  {journal} {\bibinfo
  {journal} {Phys.~Rev.~A}\ }\textbf {\bibinfo {volume} {87}},\ \bibinfo
  {pages} {022336} (\bibinfo {year} {2013})}\BibitemShut {NoStop}%
\bibitem [{\citenamefont {Bouwmeester}\ \emph {et~al.}(1999)\citenamefont
  {Bouwmeester}, \citenamefont {Marzoli}, \citenamefont {Karman}, \citenamefont
  {Schleich},\ and\ \citenamefont {Woerdman}}]{Bouwmeester1999}%
  \BibitemOpen
  \bibfield  {author} {\bibinfo {author} {\bibfnamefont {D.}~\bibnamefont
  {Bouwmeester}}, \bibinfo {author} {\bibfnamefont {I.}~\bibnamefont
  {Marzoli}}, \bibinfo {author} {\bibfnamefont {G.~P.}\ \bibnamefont {Karman}},
  \bibinfo {author} {\bibfnamefont {W.}~\bibnamefont {Schleich}}, \ and\
  \bibinfo {author} {\bibfnamefont {J.~P.}\ \bibnamefont {Woerdman}},\ }\href
  {\doibase 10.1103/physreva.61.013410} {\bibfield  {journal} {\bibinfo
  {journal} {Phys.~Rev.~A}\ }\textbf {\bibinfo {volume} {61}},\ \bibinfo
  {pages} {013410} (\bibinfo {year} {1999})}\BibitemShut {NoStop}%
\bibitem [{\citenamefont {Knight}\ \emph {et~al.}(2003)\citenamefont {Knight},
  \citenamefont {Rold\'an},\ and\ \citenamefont {Sipe}}]{Knight2003}%
  \BibitemOpen
  \bibfield  {author} {\bibinfo {author} {\bibfnamefont {P.~L.}\ \bibnamefont
  {Knight}}, \bibinfo {author} {\bibfnamefont {E.}~\bibnamefont {Rold\'an}}, \
  and\ \bibinfo {author} {\bibfnamefont {J.~E.}\ \bibnamefont {Sipe}},\ }\href
  {\doibase 10.1103/physreva.68.020301} {\bibfield  {journal} {\bibinfo
  {journal} {Phys.~Rev.~A}\ }\textbf {\bibinfo {volume} {68}},\ \bibinfo
  {pages} {020301} (\bibinfo {year} {2003})}\BibitemShut {NoStop}%
\bibitem [{\citenamefont {Goyal}\ \emph {et~al.}(2013)\citenamefont {Goyal},
  \citenamefont {Roux}, \citenamefont {Forbes},\ and\ \citenamefont
  {Konrad}}]{Goyal2013}%
  \BibitemOpen
  \bibfield  {author} {\bibinfo {author} {\bibfnamefont {S.~K.}\ \bibnamefont
  {Goyal}}, \bibinfo {author} {\bibfnamefont {F.~S.}\ \bibnamefont {Roux}},
  \bibinfo {author} {\bibfnamefont {A.}~\bibnamefont {Forbes}}, \ and\ \bibinfo
  {author} {\bibfnamefont {T.}~\bibnamefont {Konrad}},\ }\href {\doibase
  10.1103/physrevlett.110.263602} {\bibfield  {journal} {\bibinfo  {journal}
  {Phys. Rev. Lett.}\ }\textbf {\bibinfo {volume} {110}},\ \bibinfo {pages}
  {263602} (\bibinfo {year} {2013})}\BibitemShut {NoStop}%
\bibitem [{\citenamefont {Schreiber}\ \emph {et~al.}(2010)\citenamefont
  {Schreiber}, \citenamefont {Cassemiro}, \citenamefont {Poto\v{c}ek},
  \citenamefont {G\'abris}, \citenamefont {Mosley}, \citenamefont {Andersson},
  \citenamefont {Jex},\ and\ \citenamefont {Silberhorn}}]{Schreiber2010}%
  \BibitemOpen
  \bibfield  {author} {\bibinfo {author} {\bibfnamefont {A.}~\bibnamefont
  {Schreiber}}, \bibinfo {author} {\bibfnamefont {K.~N.}\ \bibnamefont
  {Cassemiro}}, \bibinfo {author} {\bibfnamefont {V.}~\bibnamefont
  {Poto\v{c}ek}}, \bibinfo {author} {\bibfnamefont {A.}~\bibnamefont
  {G\'abris}}, \bibinfo {author} {\bibfnamefont {P.~J.}\ \bibnamefont
  {Mosley}}, \bibinfo {author} {\bibfnamefont {E.}~\bibnamefont {Andersson}},
  \bibinfo {author} {\bibfnamefont {I.}~\bibnamefont {Jex}}, \ and\ \bibinfo
  {author} {\bibfnamefont {C.}~\bibnamefont {Silberhorn}},\ }\href {\doibase
  10.1103/physrevlett.104.050502} {\bibfield  {journal} {\bibinfo  {journal}
  {Phys. Rev. Lett.}\ }\textbf {\bibinfo {volume} {104}},\ \bibinfo {pages}
  {050502} (\bibinfo {year} {2010})}\BibitemShut {NoStop}%
\bibitem [{\citenamefont {Schreiber}\ \emph {et~al.}(2012)\citenamefont
  {Schreiber}, \citenamefont {G{\'a}bris}, \citenamefont {Rohde}, \citenamefont
  {Laiho}, \citenamefont {{\v{S}}tefa{\v{n}}{\'a}k}, \citenamefont
  {Poto{\v{c}}ek}, \citenamefont {Hamilton}, \citenamefont {Jex},\ and\
  \citenamefont {Silberhorn}}]{Schreiber2012}%
  \BibitemOpen
  \bibfield  {author} {\bibinfo {author} {\bibfnamefont {A.}~\bibnamefont
  {Schreiber}}, \bibinfo {author} {\bibfnamefont {A.}~\bibnamefont
  {G{\'a}bris}}, \bibinfo {author} {\bibfnamefont {P.~P.}\ \bibnamefont
  {Rohde}}, \bibinfo {author} {\bibfnamefont {K.}~\bibnamefont {Laiho}},
  \bibinfo {author} {\bibfnamefont {M.}~\bibnamefont
  {{\v{S}}tefa{\v{n}}{\'a}k}}, \bibinfo {author} {\bibfnamefont
  {V.}~\bibnamefont {Poto{\v{c}}ek}}, \bibinfo {author} {\bibfnamefont
  {C.}~\bibnamefont {Hamilton}}, \bibinfo {author} {\bibfnamefont
  {I.}~\bibnamefont {Jex}}, \ and\ \bibinfo {author} {\bibfnamefont
  {C.}~\bibnamefont {Silberhorn}},\ }\href {\doibase 10.1126/science.1218448}
  {\bibfield  {journal} {\bibinfo  {journal} {Science}\ }\textbf {\bibinfo
  {volume} {336}},\ \bibinfo {pages} {55} (\bibinfo {year} {2012})}\BibitemShut
  {NoStop}%
\bibitem [{\citenamefont {Goyal}\ \emph {et~al.}(2015)\citenamefont {Goyal},
  \citenamefont {Roux}, \citenamefont {Forbes},\ and\ \citenamefont
  {Konrad}}]{Goyal2015a}%
  \BibitemOpen
  \bibfield  {author} {\bibinfo {author} {\bibfnamefont {S.~K.}\ \bibnamefont
  {Goyal}}, \bibinfo {author} {\bibfnamefont {F.~S.}\ \bibnamefont {Roux}},
  \bibinfo {author} {\bibfnamefont {A.}~\bibnamefont {Forbes}}, \ and\ \bibinfo
  {author} {\bibfnamefont {T.}~\bibnamefont {Konrad}},\ }\href {\doibase
  10.1103/physreva.92.040302} {\bibfield  {journal} {\bibinfo  {journal} {Phys.
  Rev. A}\ }\textbf {\bibinfo {volume} {92}},\ \bibinfo {pages} {040302}
  (\bibinfo {year} {2015})}\BibitemShut {NoStop}%
\bibitem [{\citenamefont {Cardano}\ \emph {et~al.}(2015)\citenamefont
  {Cardano}, \citenamefont {Massa}, \citenamefont {Qassim}, \citenamefont
  {Karimi}, \citenamefont {Slussarenko}, \citenamefont {Paparo}, \citenamefont
  {de~Lisio}, \citenamefont {Sciarrino}, \citenamefont {Santamato},
  \citenamefont {Boyd},\ and\ \citenamefont {Marrucci}}]{Cardano2015}%
  \BibitemOpen
  \bibfield  {author} {\bibinfo {author} {\bibfnamefont {F.}~\bibnamefont
  {Cardano}}, \bibinfo {author} {\bibfnamefont {F.}~\bibnamefont {Massa}},
  \bibinfo {author} {\bibfnamefont {H.}~\bibnamefont {Qassim}}, \bibinfo
  {author} {\bibfnamefont {E.}~\bibnamefont {Karimi}}, \bibinfo {author}
  {\bibfnamefont {S.}~\bibnamefont {Slussarenko}}, \bibinfo {author}
  {\bibfnamefont {D.}~\bibnamefont {Paparo}}, \bibinfo {author} {\bibfnamefont
  {C.}~\bibnamefont {de~Lisio}}, \bibinfo {author} {\bibfnamefont
  {F.}~\bibnamefont {Sciarrino}}, \bibinfo {author} {\bibfnamefont
  {E.}~\bibnamefont {Santamato}}, \bibinfo {author} {\bibfnamefont {R.~W.}\
  \bibnamefont {Boyd}}, \ and\ \bibinfo {author} {\bibfnamefont
  {L.}~\bibnamefont {Marrucci}},\ }\href {\doibase 10.1126/sciadv.1500087}
  {\bibfield  {journal} {\bibinfo  {journal} {Sci. Adv.}\ }\textbf {\bibinfo
  {volume} {1}},\ \bibinfo {pages} {e1500087} (\bibinfo {year}
  {2015})}\BibitemShut {NoStop}%
\bibitem [{\citenamefont {Cardano}\ \emph {et~al.}(2016)\citenamefont
  {Cardano}, \citenamefont {Maffei}, \citenamefont {Massa}, \citenamefont
  {Piccirillo}, \citenamefont {de~Lisio}, \citenamefont {De~Filippis},
  \citenamefont {Cataudella}, \citenamefont {Santamato},\ and\ \citenamefont
  {Marrucci}}]{Cardano2016}%
  \BibitemOpen
  \bibfield  {author} {\bibinfo {author} {\bibfnamefont {F.}~\bibnamefont
  {Cardano}}, \bibinfo {author} {\bibfnamefont {M.}~\bibnamefont {Maffei}},
  \bibinfo {author} {\bibfnamefont {F.}~\bibnamefont {Massa}}, \bibinfo
  {author} {\bibfnamefont {B.}~\bibnamefont {Piccirillo}}, \bibinfo {author}
  {\bibfnamefont {C.}~\bibnamefont {de~Lisio}}, \bibinfo {author}
  {\bibfnamefont {G.}~\bibnamefont {De~Filippis}}, \bibinfo {author}
  {\bibfnamefont {V.}~\bibnamefont {Cataudella}}, \bibinfo {author}
  {\bibfnamefont {E.}~\bibnamefont {Santamato}}, \ and\ \bibinfo {author}
  {\bibfnamefont {L.}~\bibnamefont {Marrucci}},\ }\href {\doibase
  10.1038/ncomms11439} {\bibfield  {journal} {\bibinfo  {journal} {Nat.
  Commun.}\ }\textbf {\bibinfo {volume} {7}},\ \bibinfo {pages} {11439}
  (\bibinfo {year} {2016})}\BibitemShut {NoStop}%
\bibitem [{\citenamefont {Simon}\ and\ \citenamefont
  {Mukunda}(1989)}]{Simon1989}%
  \BibitemOpen
  \bibfield  {author} {\bibinfo {author} {\bibfnamefont {R.}~\bibnamefont
  {Simon}}\ and\ \bibinfo {author} {\bibfnamefont {N.}~\bibnamefont
  {Mukunda}},\ }\href {\doibase 10.1016/0375-9601(89)90748-2} {\bibfield
  {journal} {\bibinfo  {journal} {Phys. Lett. A}\ }\textbf {\bibinfo {volume}
  {138}},\ \bibinfo {pages} {474} (\bibinfo {year} {1989})}\BibitemShut
  {NoStop}%
\bibitem [{\citenamefont {Marrucci}\ \emph {et~al.}(2006)\citenamefont
  {Marrucci}, \citenamefont {Manzo},\ and\ \citenamefont
  {Paparo}}]{Marrucci2006}%
  \BibitemOpen
  \bibfield  {author} {\bibinfo {author} {\bibfnamefont {L.}~\bibnamefont
  {Marrucci}}, \bibinfo {author} {\bibfnamefont {C.}~\bibnamefont {Manzo}}, \
  and\ \bibinfo {author} {\bibfnamefont {D.}~\bibnamefont {Paparo}},\ }\href
  {\doibase 10.1103/physrevlett.96.163905} {\bibfield  {journal} {\bibinfo
  {journal} {Phys. Rev. Lett.}\ }\textbf {\bibinfo {volume} {96}},\ \bibinfo
  {pages} {163905} (\bibinfo {year} {2006})}\BibitemShut {NoStop}%
\bibitem [{\citenamefont {Slussarenko}\ \emph {et~al.}(2011)\citenamefont
  {Slussarenko}, \citenamefont {Murauski}, \citenamefont {Du}, \citenamefont
  {Chigrinov}, \citenamefont {Marrucci},\ and\ \citenamefont
  {Santamato}}]{Slussarenko2011}%
  \BibitemOpen
  \bibfield  {author} {\bibinfo {author} {\bibfnamefont {S.}~\bibnamefont
  {Slussarenko}}, \bibinfo {author} {\bibfnamefont {A.}~\bibnamefont
  {Murauski}}, \bibinfo {author} {\bibfnamefont {T.}~\bibnamefont {Du}},
  \bibinfo {author} {\bibfnamefont {V.}~\bibnamefont {Chigrinov}}, \bibinfo
  {author} {\bibfnamefont {L.}~\bibnamefont {Marrucci}}, \ and\ \bibinfo
  {author} {\bibfnamefont {E.}~\bibnamefont {Santamato}},\ }\href {\doibase
  10.1364/oe.19.004085} {\bibfield  {journal} {\bibinfo  {journal} {Opt.
  Express}\ }\textbf {\bibinfo {volume} {19}},\ \bibinfo {pages} {4085}
  (\bibinfo {year} {2011})}\BibitemShut {NoStop}%
\bibitem [{\citenamefont {Bliokh}\ \emph {et~al.}(2011)\citenamefont {Bliokh},
  \citenamefont {Ostrovskaya}, \citenamefont {Alonso}, \citenamefont
  {Rodr{\'\i}guez-Herrera}, \citenamefont {Lara},\ and\ \citenamefont
  {Dainty}}]{Bliokh2011}%
  \BibitemOpen
  \bibfield  {author} {\bibinfo {author} {\bibfnamefont {K.~Y.}\ \bibnamefont
  {Bliokh}}, \bibinfo {author} {\bibfnamefont {E.~A.}\ \bibnamefont
  {Ostrovskaya}}, \bibinfo {author} {\bibfnamefont {M.~A.}\ \bibnamefont
  {Alonso}}, \bibinfo {author} {\bibfnamefont {O.~G.}\ \bibnamefont
  {Rodr{\'\i}guez-Herrera}}, \bibinfo {author} {\bibfnamefont {D.}~\bibnamefont
  {Lara}}, \ and\ \bibinfo {author} {\bibfnamefont {C.}~\bibnamefont
  {Dainty}},\ }\href {\doibase 10.1364/oe.19.026132} {\bibfield  {journal}
  {\bibinfo  {journal} {Opt. Express}\ }\textbf {\bibinfo {volume} {19}},\
  \bibinfo {pages} {26132} (\bibinfo {year} {2011})}\BibitemShut {NoStop}%
\bibitem [{\citenamefont {Piccirillo}\ \emph {et~al.}(2010)\citenamefont
  {Piccirillo}, \citenamefont {D'Ambrosio}, \citenamefont {Slussarenko},
  \citenamefont {Marrucci},\ and\ \citenamefont {Santamato}}]{Piccirillo2010}%
  \BibitemOpen
  \bibfield  {author} {\bibinfo {author} {\bibfnamefont {B.}~\bibnamefont
  {Piccirillo}}, \bibinfo {author} {\bibfnamefont {V.}~\bibnamefont
  {D'Ambrosio}}, \bibinfo {author} {\bibfnamefont {S.}~\bibnamefont
  {Slussarenko}}, \bibinfo {author} {\bibfnamefont {L.}~\bibnamefont
  {Marrucci}}, \ and\ \bibinfo {author} {\bibfnamefont {E.}~\bibnamefont
  {Santamato}},\ }\href {\doibase 10.1063/1.3527083} {\bibfield  {journal}
  {\bibinfo  {journal} {Appl. Phys. Lett.}\ }\textbf {\bibinfo {volume} {97}},\
  \bibinfo {pages} {241104} (\bibinfo {year} {2010})}\BibitemShut {NoStop}%
\bibitem [{\citenamefont {Padgett}\ and\ \citenamefont
  {Allen}(1995)}]{Padgett1995}%
  \BibitemOpen
  \bibfield  {author} {\bibinfo {author} {\bibfnamefont {M.}~\bibnamefont
  {Padgett}}\ and\ \bibinfo {author} {\bibfnamefont {L.}~\bibnamefont
  {Allen}},\ }\href@noop {} {\bibfield  {journal} {\bibinfo  {journal} {Opt.
  Commun.}\ }\textbf {\bibinfo {volume} {121}},\ \bibinfo {pages} {36}
  (\bibinfo {year} {1995})}\BibitemShut {NoStop}%
\bibitem [{\citenamefont {Horn}\ and\ \citenamefont
  {Johnson}(2012)}]{Hornmatrix}%
  \BibitemOpen
  \bibfield  {author} {\bibinfo {author} {\bibfnamefont {R.~A.}\ \bibnamefont
  {Horn}}\ and\ \bibinfo {author} {\bibfnamefont {C.~R.}\ \bibnamefont
  {Johnson}},\ }\href@noop {} {\emph {\bibinfo {title} {Matrix analysis}}}\
  (\bibinfo  {publisher} {Cambridge university press, New York},\ \bibinfo
  {year} {2012})\BibitemShut {NoStop}%
\end{thebibliography}

%

\end{document}